\documentclass[runningheads]{llncs}

\usepackage[utf8x]{inputenc}
\usepackage{graphicx}
\usepackage{multirow}
\usepackage{subfigure}
\usepackage{epstopdf}
\usepackage{color}
\usepackage{enumitem}
\usepackage{amsmath}

\newcommand{\no}[1]{}

\newcommand{\acumM}{\mathsf{AcumM}}                     
\newcommand{\acumMname}{\mathsf{Accumulated-Matrix }}   
\newcommand{\tctr}{\mathsf{TTCTR}}     
\newcommand{\tctrname}{\mathsf{Topology\&Trip-aware~CTR}}     

\newcommand{\ctr}{\mathsf{CTR}}       

\newcommand{\csa}{\mathsf{CSA}}
\newcommand{\icsa}{\mathsf{iCSA}}

\newcommand{\wm}{\mathsf{WM}}

\begin{document}
\authorrunning{N. Brisaboa, A. Fari\~na, D. Galaktionov, T. Rodeiro, and M.A. Rodr\'iguez}

\titlerunning{New structures to solve aggregated queries for trips over pub. transp. netws.}

\title{New structures to solve aggregated queries for trips over public transportation 
	networks%
		\thanks{
			\footnotesize{
				Funded in part by European Union's Horizon 2020 research and innovation programme
				under the Marie Sklodowska-Curie grant agreement No 690941 (project BIRDS).
				The Spanish group is also partially funded
				by Xunta de Galicia/FEDER-UE [CSI: ED431G/01 and GRC: ED431C 2017/58]; 
				by MINECO-AEI/FEDER-UE [Datos 4.0: TIN2016-78011-C4-1-R; 
				Velocity: TIN2016-77158-C4-3-R; and ETOME-RDFD3: TIN2015-69951-R]; 
				by MINECO-CDTI/FEDER-UE [INNTERCONECTA: uForest ITC-20161074]; and
                by FPI Program [BES-C-2017-0085].
				M. A. Rodr\'iguez  is partially funded by Fondecyt-Conicyt grant number 1170497 and by the Millennium Institute for Foundational Research on Data. 	}
		}
}

\author{Nieves R. Brisaboa \inst{1} \and
	Antonio Fari\~na \inst{1} \and 
	Daniil Galaktionov\inst{1}\inst{2} \and \\
	Tirso V. Rodeiro\inst{1} \and
	M. Andrea Rodr\'iguez\inst{3}\inst{4} 
}

\institute{Universidade da Coru\~na, Fac. Inform\'atica, CITIC, Spain
\and
Enxenio S.L., Spain
\and  
	Universidad de Concepcion, Computer Science Department, Chile 
\and Millennium Institute for Foundational Research on Data, Chile
}

\maketitle


\begin{abstract}
	Representing the trajectories of mobile objects is a hot topic from the widespread use of smartphones and other GPS devices. However, few works have focused on representing 
	trips over public transportation networks (buses, subway, and trains) where user's trips can be seen as a sequence of stages performed within a vehicle shared with many other users. In this context, representing vehicle journeys reduces the redundancy because all the passengers inside a vehicle share the same arrival time for each stop. 
	In addition, each vehicle journey follows exactly the sequence of stops corresponding to its 
	line, which makes it unnecessary to represent that sequence for each journey.

	
	To solve data management for transportation systems, we designed a conceptual model that gave us a better insight into this data domain and allowed us the definition of relevant terms and the detection of redundancy sources among those data. Then, we designed two compact representations focused on users' trips ($\tctr$) and on vehicle trips  ($\acumM$), respectively. Each approach owns some strengths and is able to answer some queries efficiently.
	
	

	We include experimental results over synthetic trips generated from accurate schedules obtained from a real network description (from the bus transportation system of Madrid) to show the space/time trade-off of both approaches. 
	We considered a wide range of different queries 
	about the use of the transportation network such as counting-based/aggregate queries regarding the load of any line of the network at different times.  

\end{abstract}

\section{Introduction}
The management of public transportation systems is a complex problem that has been typically faced from the point of view of the {\bf offer} (lines, stops, schedules of journeys for each line, ...).
In the last decade, the widespread use of new technologies allowing somehow the tracking of users' movements along a network transportation system (mobile phones with GPS, use of RFID technologies, smart cards used to pay and enter buses/trains, ...) brings new opportunities to gather the actual usage of the transportation systems allowing to study the problem from the point of view of  {\bf users' demand}. In consequence, it is now possible to develop new applications to exploit those data in order to effectively handle the resources of the transportation system and to give a better service to the users.

The management of the transportation system has become a Big Data problem in many important cities around the world, where millions of passengers use the public transportation network every day.
Therefore, even though we can assume the gathered data is reliable (even in the case of depending on the smart cards provided to users that typically gather only the entry point to the network,  the end point can commonly be derived using historical data from user trips and transportation models~\cite{Munizaga20129}), the problem lies now on how to represent user trips in such a way that not only we provide a compact representation but also we enable performing queries in an efficient way.

While there exist many works that tackle the problem of representing trajectories of mobile objects constrained to networks~\cite{DBLP:journals/josis/RichterSL12,DBLP:journals/jss/KellarisPT13,DBLP:conf/w2gis/FunkeSSS15}, they typically aim at locating the position of those objects from the underlying trajectories. Others ~\cite{DBLP:conf/gis/KroghPTT14,DBLP:conf/gis/KoideTY15} focus on solving {\em strict and approximate path queries} that permit to find the trajectories that follow a given line pattern within a given time interval. The latter work \cite{DBLP:conf/gis/KoideTY15} is, to the best of our knowledge, the first work using a compact data structure to represent the spatial data (a  FM-index~\cite{DBLP:conf/focs/FerraginaM00}).  Yet, none of them have been designed to tackle the analysis of the usage of the transportation network and would hardly support queries such as {\em count the number of 
	user trips that went from stop $X$ to stop $Y$}, or {\em show the load of the lines at a given hour}.

In \cite{BRISABOA20181}, a representation for user trips along a transportation network referred to as $\ctr$ was presented. 
The different stops from bus 
lines were given 
a \textit{node-ID}. Then, 
each user trip 
was associated a string composed of the sequence of \textit{node-IDs} traversed. 
Finally, a $\csa$-based representation \cite{Sad03} was used to represent the collection with all users' trips, and a Wavelet Matrix ($\wm$) \cite{CNO15} aligned with the $\csa$ 
represented, for each trip, the time instant when 
every node from that trip was reached. $\ctr$ enabled answering counting-based aggregated queries (number of users that started/ended a trip at a given node within a given time interval, number of users that used a 
node, top-k most used nodes, etc.). In addition, since it represented the actual trips in a compact self-indexed way, $\ctr$ still possessed enough flexibility to support more complex queries. 
$\ctr$ succeeded at providing a compact representation for general trips. Yet, it  still represented trips in a redundant way when considering public transportation by bus, train, and subway. This happens because it does not exploit the fact that all the passengers in the same bus/train traverse each stop at the same time, nor the topology of the network (for all the users' trips from a stop $X$ to a stop $Y$ along a given line, all the intermediate stops are always the same).

In this work, we have analyzed the problem of representing both \textit{offer} (stops, lines, schedule for each line) and \textit{users' demand} (user trips, and stages that include stops where users get on/off or switch lines) within a public transportation network.
We present a conceptual model that 
provides valuable insights into this domain and shows both the data needed and the relationships among them.

Then, we present two complementary structures to represent those data and show how they handle some useful queries in this context.
The first solution is named $\tctrname$ ($\tctr$) and is based on a modification of $\ctr$ that also represents all the user trips but exploits both the network topology and the fact that all the passengers of the same vehicle journey reach the same stops at the same time, hence temporal information can be related to the vehicle journey rather than to each user trip. 
Therefore, it still makes up a general representation focused on users' trips. The second solution, named $\acumMname$ ($\acumM$), does not actually represents user's trips. 
It focuses on the journeys of each line, and 
accumulates the number of passengers that get on/off in each stop of each journey. Therefore, $\acumM$ is a summarization of the load each line had considering each of its journeys, in the same way a data warehouse is a summarization of the operational data in a database. 

The structure of the paper is as follows. In Section~\ref{sec:modeling} we discuss the conceptual model associated to the network transportation problem and provide some definitions. In Section~\ref{sec:common}, we present our representation of the {\em offer} (lines, stops and journeys) which is then used in our two solutions. The next sections describe both $\tctr$ and $\acumM$ and discuss the types of queries they are designed for. 
Section~\ref{sec:experiments} includes experimental results to show the space/time trade-offs of our proposals. And finally, conclusions and future work are discussed in Section~\ref{sec:conclusions}.

\section{A model to describe a public transportation network} \label{sec:modeling}

The E-R conceptual model at Figure~\ref{fig:er} represents the relevant data of any public transportation system including data related with both the {\em offer} and the {\em demand}. We did not include entities such as vehicles or drivers, as they are out of the scope of this work.
To create that model we have defined the following concepts:

\begin{itemize}
	\item  \textbf{Stop (or Stop-place)}. Places were passengers can get on/off from a vehicle. 
	
	\item \textbf{Lines}. A line (or route) is a sequence of stops that starts at a given stop $X$ and ends in another stop $Y$. 
	We consider a line 
	and its {\em return} line 
	as different lines because they include different sequences of stops.
	
	\item \textbf{Journeys}.  
	We define a {journey} (or vehicle journey, or line trip) as a trip that a vehicle performs. It departs at a given day and time from the first stop of a specific line and follows the complete sequence of stops of that line until the ending stop, allowing passengers to get on and to get off in each stop. For instance, a \textit{journey} is the trip that a bus performed along line $L1$  departing at 9:00am on 2017/05/05, and stopping at each stop of the line. 
	
	In addition to the day and time each journey starts, it could be interesting to have the time at which each stop was reached by each specific journey of the line. Yet, if such an accurate time is not needed, we can save a large amount of space by only storing, for each stop of a line, the average accumulated time needed to reach that stop from the initial stop (as in the examples shown along this paper). Some other solutions,  with different trade-off between accuracy of the temporal data and storage space, are possible. For example, we could store the average time to get to each stop of a line in peak and non-peak hours. In any case, all those strategies enable us to associate temporal data to the user-trips done within a given vehicle journey.
	


	\begin{figure}[]
		\vspace{-0.2cm}
		\begin{center}			{\includegraphics[width=0.95\textwidth]{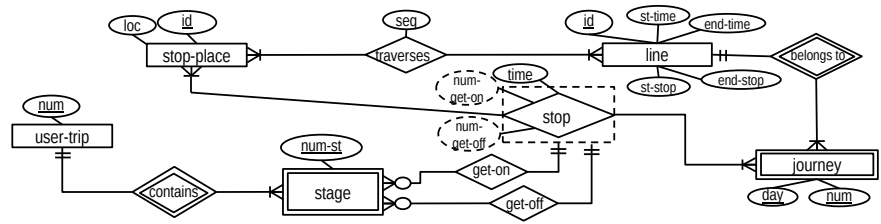}}
		\end{center}
		\vspace{-0.3cm}
		\caption{E-R model for a public transportation network.}
		\label{fig:er}
		\vspace{-0.2cm}
	\end{figure}
	
	\item \textbf{Stages}. A stage represents the pair of stops where a given user respectively gets on/off to/from a vehicle doing a given journey of a given line. 
	
	\item  \textbf{User\_trip}. We define a user trip as a sequence of stages. That is, a user trip can begin at stop $A$ from line $3$ and continue up to stop $B$ (first stage), then change to line $2$ up to stop $C$ (second stage), and so on. This enables tracking user trips from an origin to a destination. 
	Note that, since stages refer to a given journey, and we can know the time when a journey traverses a stop $X$, we can also know when a given user trip reached such stop $X$. 
	
	
	
\end{itemize}

\vspace{-6mm}
\section{Towards a practical representation: common structures to represent the {\em offer}} \label{sec:common}


In Sections \ref{sec:tctr} and \ref{sec:accumM}, we present two 
representations. The first one is focused on the representation of user trips, whereas the second one is focused on the journeys of each line and basically stores the number of users that get on/off at any given stop for each journey of a any line.
Both techniques require some common structures that handle the data that represents the {\em offer} of public transportation the network provides. Such {\em offer} refers to the structure of the network and includes the representation of the lines, and, for each line, the schedule of its journeys; that is,   their departure time from the starting stop. In addition, we use two aligned arrays for each line, one with the sequence of stops, and another with the average accumulated time to reach each stop from the first stop of the line.
Note that instead of assigning a unique sequence of estimated times to reach any stop from the line, we  could have dealt 
with several, probably more accurate, estimations for peak/low periods, or even we could have stored the actual time 
each journey reached each stop. In any case, we can estimate the time when each journey reaches each stop.


\begin{figure}[ht]
	\vspace{-0.2cm}
	\begin{center}
		{\includegraphics[width=1.00\textwidth]{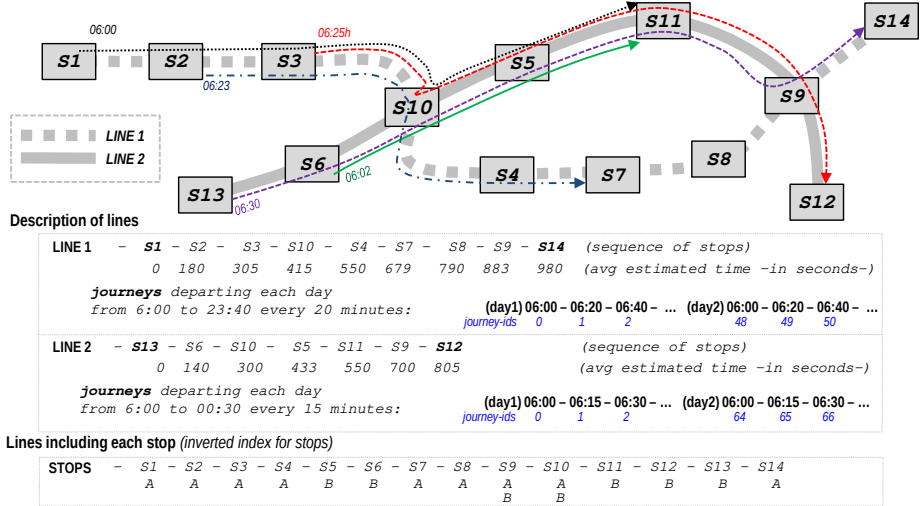}}
	\end{center}
	\vspace{-0.3cm}
	\caption{Example of (bus) public transportation network. }
	\label{fig:network}
	\vspace{-0.2cm}
\end{figure}

Figure~\ref{fig:network} includes an example of a bus network with two lines ($1$ and $2$). For each line we show the stops that compose it (e.g. Line $2$ contains the sequence of  stops $\langle S13, S6, S10, S5, S11, S9, S14 \rangle$) and the accumulated times from the initial stop (e.g. the average time to reach the fourth stop $S5$ from the starting stop of the line is $433$ seconds). We also include the starting times for each journey of each line. In this case, Line $1$ has $48$ journeys per day, the first one starts at 6:00am, the second one at 6:20am, etc. 

Note that given a line $X$ we have direct access to the information related to the $i$-th stop. Yet, 
given a stop, we do not know the line/s it belongs to. To overcome this issue, we include, for each stop $Y$, the list of lines that include such stop $Y$. It is referred to as {\em inverted index for stops} in the bottom of Figure~\ref{fig:network}. 

To sum up, we saw that to represent the network offer 
we need: {\em (i)} a sequence of stops for each line\footnote{We also store average estimated times to reach each stop from first stop of the line.}; {\em (ii)} a schedule with the starting times of the journeys of each line;
and {\em (iii)}, an inverted index to mark the lines each stop belongs to.

\medskip
Apart from the network offer, in Figure~\ref{fig:network} we also include (arrows) five user trips done along the network. 
For example, there is a user trip (dashed arrow from $S3$) that starts at stop $S3$ at 06:25am on {\em day-1} (6:20am + 305 sec.), and follows the journey of line $1$ until $S10$, where the user switches to line $2$ at time 6:35am (6:30am + 300 sec.) and continues the corresponding journey of line $2$ (the one started as 06:30h in $S13$) up to stop $S12$. That is, it includes two stages.

\section{$\tctrname$ ($\tctr$): a representation focused on user's trips } \label{sec:tctr}

A previous representation for user trips along a transportation network, named $\ctr$ \cite{BRISABOA20181}, basically associates an integer $s_i$ to each stop in the network, and represents a user trip $t_i$ as the sequence of the stops traversed plus a terminator $\$$ ($t_i= s_1,s_2,\dots,s_k\$$). Finally, a $\csa$-based representation is used to represent the collection with all users' trips, and a Wavelet Matrix ($\wm$) aligned with the $\csa$ keeps, for each trip, the (discretized) time instant in which every stop from that trip was reached. For example, the trip from Figure~\ref{fig:network} that started at $S3$ would be represented as $\langle S3,S10,S5,S11,S9,S12,\$\rangle $ and the times associated to those stops would be discretized into $5$-min time periods. $\ctr$ exploited the indexing capabilities of the underlying $\csa$ and $\wm$ to solve counting-based spatial, temporal, and spatio-temporal queries. 

Our  $\tctr$ is an adaptation of $\ctr$ that represents a user trip as a sequence of stages rather than as a sequence of stops (hence exploiting the topology of the network). 
Furthermore, instead of having to represent the time each user trip reaches a stop, we will only store a reference/$id$ of the journey (within a vehicle of a line) that the user used. The building process of $\tctr$ is presented below.

Let us assume a network with $n_s$ stops ($S$) numbered $s \in [1, n_s]$; $n_l$ lines ($L$) numbered $[1,n_l]$, and that there are $n_j^l$ journeys ($J^{l}$) for each line $l \in L$  numbered $[0, n_j^l -1]$. 
Additionally, we have the starting times for each journey and the accumulated average times for the stops of each line as discussed in the previous section.
We can define that a user gets on/off from a vehicle following the journey $j$ of line $l$ at a given stop $s$, as a triple $(s,l,j)$ where $~ l \in L,~ s \in S,~ j \in J^l$. 


Let us define $\mathcal{T} = \{t_1, t_2,\dots,  t_z \}$ as a set of $z$ user trips. Since we want to represent a user trip $t_x$ as a sequence of $k$ stages, but it holds that the final stop of a stage and the starting stop of the next stage are the same (or close in walking distance), it is not necessary to explicitly represent the final stop of each stage, except for the final stop. We define $t_x = \langle(s_1,l_1,j_1),(s_2,l_2,j_2), \dots,(s_{k+1},l_{k+1},j_{k+1}) \rangle$, $k\geq 1$. That is, we have a sequence of $k$ triples that indicate that the user got on a vehicle corresponding to the $j_i$-th journey of line $l_i$ at stop $s_i$. The last triple indicates where the user finally got off. Note that, for the last two triples, $l_k = l_{k+1}$, and $j_k = j_{k+1}$ since the beginning of the last stage is represented by the $k$-th triple, and its end by the $(k+1)$-th triple.

\begin{example}
	Assuming that all the user trips depicted in Figure~\ref{fig:network} belong to our 1st-day, 
	those user trips can be represented as: 
	$t_1 = \langle (1,1,\textbf{0}), (10,2,\textbf{1}), (11,2,\textbf{1})  \rangle$, 
	$t_2 = \langle (2,1,\textbf{1}), (7,1,\textbf{1}) \rangle$, 
	$t_3 = \langle (3,1,\textbf{1}), (10,2,\textbf{2}), (12,2,\textbf{2}) \rangle$, 
	$t_4 = \langle (6,2,\textbf{0}),(11,2,\textbf{0}) \rangle$, and 
	$t_5 = \langle (13,2,\textbf{2}), (9,1,\textbf{2}), (14,1,\textbf{2}) \rangle$.  
	Note that, for example, $(13,2,\textbf{2})$ from $t_5$ indicates that, at stop $13$, the user got on a vehicle from line $2$, that corresponds to the $\textbf{2}$-nd journey. We know it is the \textbf{2}-nd journey because  $t_5$ started at 06:30h, which is the departure time of journey \textbf{2}. Note also that the line and journey $ids$ of the last triple of each trip are identical to the ones in the previous triple.
	\qed	\label{ex:trips}
\end{example}

In $\tctr$, we represent both the spatial (lines and stops) and the temporal component (journeys) of the user trips of our collection of trips $\mathcal{T}$ using respectively a $\csa$ and a $\wm$ aligned with the $\csa$. In the following sections we show how we handle such components, and how we solve some queries of interest.

\subsection{Representing the spatial component of $\tctr$ with a $\csa$}
\label{sec:csa}
We use a variant of the $\csa$  \cite{FBNCPR12} for integer alphabets to represent the spatial component, i.e. the sequence of pairs $(s_i,l_i)$ that compose each user trip in $\mathcal{T}$. However, in order to create a $\csa$ we need to assign each pair $(stop,line)$ a unique integer $id$. This will allow us to create an integer sequence $S[1,n]$ (ended by a $\$$  terminator) over which our $\csa$ will be built. For this purpose we create a vocabulary $V$ (with $1+n_s(1+n_l)$ entries)  as follows:
\begin{itemize}
	\item Entry $V[0]$ is reserved for the terminator symbol $\$$.
	\item Entries $\langle V[1],V[2], \dots V[n_s]\rangle]$ are associated to stops $\langle 1,2,\dots, n_s\rangle$ and are used to represent the final stops of the trips. That is, when a given stop $x$ ends a user trip, it is given $id \leftarrow x$.
	\item The last $n_l$$\times$$n_s$ entries are associated to the sequence composed of the pairs $(s,l) \in S$$\times$$L$ considering that those pairs are sorted by $s$ and $l$ respectively. That is, entry $V[n_s+1]$ is given to $(s,l)=(1,1)$; $V[n_s+2]$ to $(1, 2)$; $V[n_s+3]$ to $(1, 3)$; $\dots$; $V[n_s+n_l]$ to $(1, {n_l})$;  $V[n_s+n_l +1]$ to $(s,l)= (2, 1)$, $V[n_s+n_l +2]$ to $(2, 2)$, and so on. In practice, in this case, the $id/${\em pos-in-V} for a pair $(s,l)$ is obtained as $id \leftarrow  n_s+ n_l(s-1) + l$. 
\end{itemize}

Note that there will be a large number of unused entries (holes) in $V$. Yet, this can be efficiently handled by a bitvector $B$ with rank/select capabilities 
that marks the used entries from $V$. Therefore, once we gather the position ($id$) corresponding to a pair $(s,l)$ in $V$, we obtain its final position ($id'$) in a {\em vocabulary without holes} ($V'$) as $id' \leftarrow rank_1(B,id)$. Our $id'$ assignment ensures that the used pairs $(s_i,l)$, corresponding to a given stop $s_i$, receive contiguous $id'$s. This will be interesting at query time.  

The next step processes each user trip $t_i \in \mathcal{T},~i \in [1,z]$ replacing all the pairs $(stop,line)$ in $t_i$ (except in the last one where the line is already known and we only need $s$)  by their corresponding $id'$. After each trip, a $0$ ($id'$ for terminator $\$$) is added. That is, a trip $t_i$ with $k$ stages is regarded as a string $s_1s_2\dots s_{k} s_{last} \$$, where $1\leq s_{last} \leq rank_1(B,n_s)$ is the $id'$ of ending stop given to stop $s$. 

\begin{figure}[tbhp]
	\vspace{-0.2cm}
	\begin{center}
		{\includegraphics[width=1.00\textwidth]{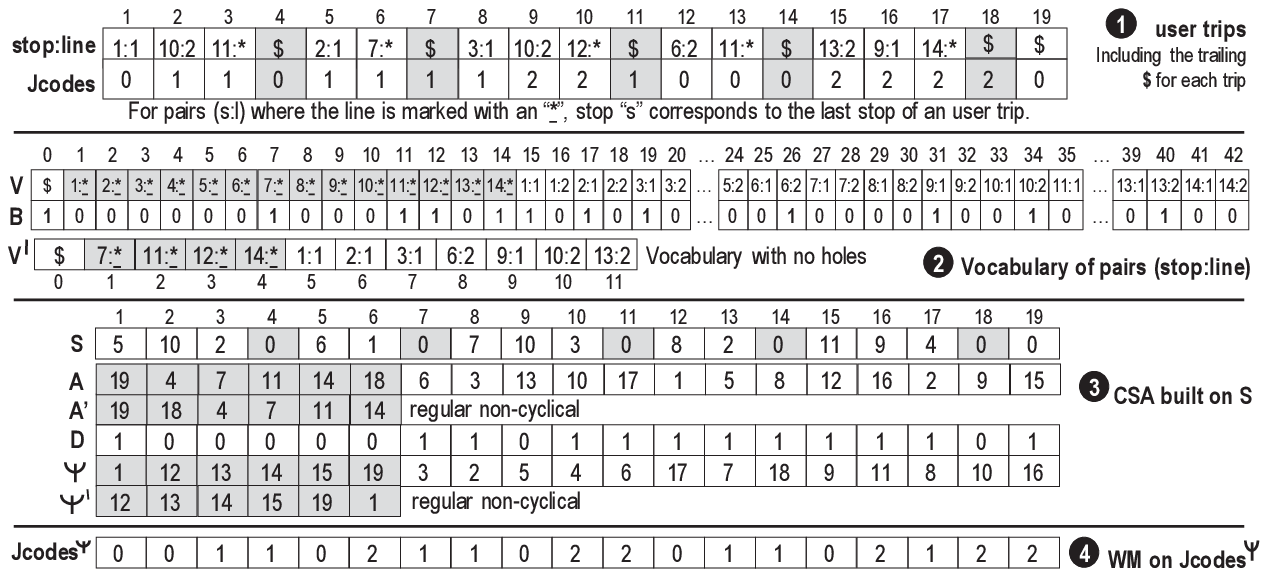}}
	\end{center}
	\vspace{-0.3cm}
	\caption{Structures involved in the creation of a $\tctr$. }
	\label{fig:ttctr}
	\vspace{-0.2cm}
\end{figure}

Once this process has completed, a sequence $S[1,n]$ that contains only values from $V'$ is obtained, and a $\csa$ can be built on $S$. 

In parallel with the construction of $S$, we create a sequence $Jcodes[1,n]$ aligned to $S$ where we set, for each trip $t_i$, the $journey$-$id$ corresponding to each stage in $t_i$. Recall the $journey$-$id$ is the third term from the triples $(s_i,l_i,j_i),~i\in[1,k]$ from $t_i$. In addition, assuming that $S[p]$ contains the $0$ corresponding to the terminator $\$$ for the trip, we set $Jcodes[p]\leftarrow j_1$ (i.e. the same journey-id as  the starting stop of the trip). According to the discussion above, Figure~\ref{fig:ttctr} shows, for the user trips in Example~\ref{ex:trips}: {\em (step-1)} the sequence of pairs $(stop,line)$ for each trip, and the corresponding $Jcodes$; {\em (step-2)} the vocabularies, including $V$, $B$, and $V'$ (ending stops $s$ of trips do not need the line, therefore we use ``$s\!:\!*$".);\footnote{In this example, with only $5$ trips,  we have only $11$ {\em used pairs} in $V$, but in a real scenario for each stop of each line ({\em existing pair} $(s,l)$) 
	there will be a $1$ in $B$.} 
and {\em (step-3)} the structures involved in the creation of $\csa$ from which $\tctr$ uses $\Psi$, $D$, (and $V'$, $B$).


As in $\ctr$ \cite{BRISABOA20181}, we sort the terminators considering that each trip is a cyclical string. For instance in Figure~\ref{fig:ttctr}, $S[18]=0$ would traditionally be followed by $S[19]=0$, but for sorting purposes we consider it is instead followed by $S[15]=11$. After that, we make $\Psi$ cyclical in the terms of each user trip. That is, let us assume that a user trip lays on $S[i] \dots S[i+k+1]$, i.e. the terminator of the trip is at position $e=i+k+1$ in $S$. Therefore, being $A[j]=e$, we modify  $\Psi[j]$ in such a way that $A[\Psi[j]]$ points not to the initial position $e+1$ of the next user trip, but it cyclically points to the beginning of the same trip; that is, we set $\Psi[j]\leftarrow \Psi[A^{-1}[i]]$.\footnote{Note that $A^{-1}$ is the inverse of the suffix array $A$; i.e. $A^{-1}[i]=j$ iff $A[j]=i$.} 
Using a cyclical $\Psi$ will enable searching efficiently for user trips that started at a stop $X$ and ended at a stop $Y$, as we will see below.





\subsection{Representing the temporal component of $\tctr$ with a  $\wm$}

The temporal component of $\tctr$ includes the sequence $Jcodes$ described above. Recall $Jcodes$ contains $journey$-$ids$ aligned to the values in $S$, and that, for every line $l$ there are $n_j^l$ journeys sorted by their starting time and numbered as $0\dots n_j^l-1$, and also we have average accumulated times to reach each stop in the line. Therefore, this representation allows us to describe exact times for each stop. 
In practice, we use $Jcodes^{\Psi}$, which is aligned to $\Psi$ rather than to $S$. See {\em step-4} in Figure~\ref{fig:ttctr}. 
Note that $Jcodes[8]= 1$ corresponds to $Jcodes^{\Psi}[14]=1$, since $A[14]=8$; 
          $Jcodes[9]= 2$ corresponds to $Jcodes^{\Psi}[18]=2$, since $A[18]=9$; and so on.
We represent the sequence $Jcodes^{\Psi}$ with a $\wm$. This saves space, and provides indexing capabilities to the temporal component. 

\section{Dealing with aggregated data: $\acumMname$} \label{sec:accumM}

We propose $\acumM$ as an intuitive solution for the representation of two-dimensional matrices of integers 
with support to aggregated queries by row, column, or window/range. In the context of a public transportation network we found  queries referred to a given line where data must be aggregated either by {\em stop} (e.g. number of users that got on a vehicle at stop $X$); by {\em time-interval}, hence referred to a sequence of consecutive journeys within that time-interval (e.g. number of users got on at any stop of the line on 2017/03/24); or by {\em stop and time-interval}.

\begin{figure}[th]
	\vspace{-0.2cm}
	\begin{center}
		{\includegraphics[width=1.0\textwidth]{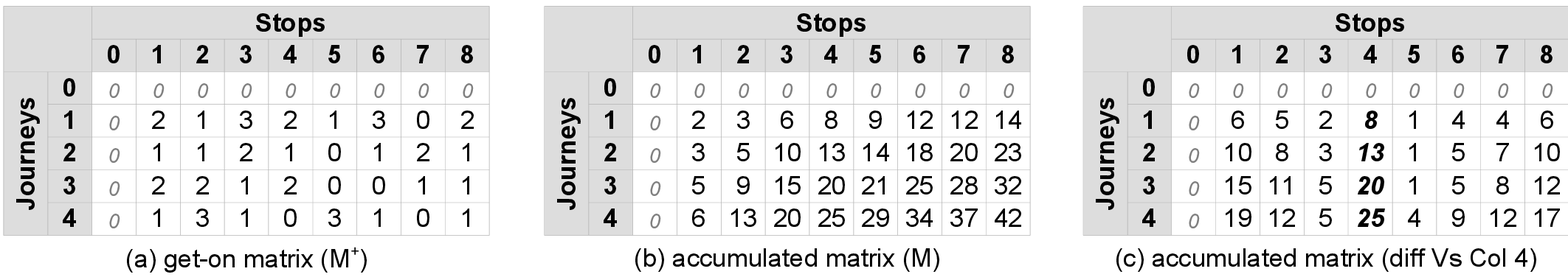}}
	\end{center}
	\vspace{-0.3cm}
	\caption{Example, \textbf{for a given line}, of get-on matrix (a), accumulated matrix (b), and compact representation with gaps (c). Stop and column $\mathbf{0}$ are virtual. The values in column/row $\mathbf{0}$ are set to $zero$ to simplify operations. }	
	\label{fig:accM}
	\vspace{-0.2cm}
\end{figure}

Let us assume that, for a given line, we have a matrix $M^+$  that stores the number of users that got-on  at each stop (column) from each journey (row). Figure~\ref{fig:accM}.(a) includes an example. 
To efficiently support aggregated queries, we create, for each line, the accumulated get-on matrix $M$ for $M^+$.  
We compute the value of a cell $M(r,c) \leftarrow \sum_{i=1}^r \sum_{j=1}^c M^+(i,j)$. That is, each cell contains the sum of all the values from position $(1,1)$ to position $(r,c)$.  
$M$ is depicted in Figure~\ref{fig:accM}.(b). 
The accumulated matrix $M$ allows us to solve a range count query over $M^+$ in $O(1)$ time by computing:~ $\mathsf{countRange((x_1,y_1),(x_2,y_2))} \leftarrow {M(x_2,y_2)} - {M(x_2,y_1-1)} - {M(x_1-1,y_2)} + {M(x_1-1,y_1-1)}$.
In $\acumM$, we actually represent, for each line, two accumulated matrices that  count, respectively, the passengers that get on and get off to/from a journey in each stop.

$\mathsf{countRange}$ allows us to add: {\em (i)} consecutive values of a column (e.g. users that got on in a stop $X$ in consecutive journeys, such as those in one day); {\em (ii)} consecutive values in a row (e.g. users that, for a given journey, got on along a consecutive sequence of stops, such as those in a neighborhood); and {\em (iii)} values in a window (e.g. users that got on in a consecutive sequence of stops in consecutive journeys).



Being $C$ the capacity of a vehicle, a simple way to decrease space usage on $M$ (it has $n_s$$\times$$ n^l_j$ integers) consists in keeping the middle column $m\leftarrow (n_s+1)/2$ explicitly, and representing the values in the other columns $m\pm k$ as the difference with respect to column $m$. This is depicted in Figure~\ref{fig:accM}.(c). The differences in columns $m \pm k$ require at most $\lceil \log_2 kC \rceil$ bits, while retaining direct access.

%
%


\vspace{-3mm}
\section{Performing queries on $\tctr$ and $\acumM$} \label{sec:queries}

$\acumM$ and $\tctr$ are designed for different purposes and therefore each highlight in different types of queries.


\subsubsection{Queries for $\acumMname$: } 
This data structure resembles a data-warehouse, i.e. 
it stores aggregated values (rather than individual trips) 
to  efficiently answer aggregated queries about the number of users (load of the network)
in a given stop or group of  stops over one or more journeys. 
Recall $\mathsf{countRange}$  efficiently 
sums the values of any submatrix. Yet, some useful queries could need more than one $\mathsf{countRange}$ operation, and then to either aggregate or compute the average of those results. For example, if we want to know the {\em average number of users that got on in line $L$ in stops of a neighborhood (consecutive stops) between 8:00 and 9:00 along the last month}, we will need a $\mathsf{countRange}$ operation for each window including the consecutive stops and the consecutive journeys inside that period for each day. Finally, we 
add the results of those $\mathsf{countRange}$ operations (one per day) and divide the result by the number of days in the month.

Since we have both the accumulated matrix for users getting on and getting off, we can easily compute queries about the load of the journeys. For example, to know \textit{how many users were inside the vehicle of journey $j$ from line $L$, between the stops $X$ and $X$$+$$1$}, we 
compute: {$tot\_up \leftarrow$}\textit{ how many users got on in such journey $j$ between stops $1$ and $X$} (inclusive), using the accumulated get-on matrix for line $L$ ($\mathsf{countRange}$ of a row);  in the same way, using the accumulated get-off matrix, we compute $tot\_down \leftarrow$\textit{how many users got off in the same journey and range of stops}; and finally,  we return the value $total \leftarrow tot\_up - tot\_down$.

\subsubsection{Queries for $\tctr$:} 

Recall that $\tctr$ actually stores all the individual trips. This allows  it to answer queries about the patterns users follow when using the transportation network. For example, queries about \textit{how many users start their trips in stop $X$}, or \textit{end their trips in stop $Y$}, or even\textit{ started their trips in stop $X$ and ended in stop $Y$}, can be efficiently answered because  $\csa$ easily locates the subsection devoted to each stop, and the cyclic encoding of the trips allows to ask for patters such as $\$X$ or $Y\$$ or even $Y\$X$. Note also that our way to encode the pairs $stop$:$line$ guaranties that the occurrences of a stop for different lines are consecutive in the $\csa$, therefore, we can ask both \textit{how many users start their trips in stop $X$} or \textit{how many users start their trips in stop $X$ of line $L$}.
Finally, using the $WM$ we can filter out those queries by time using the journeys.

Note that none of those queries can be answered by $\acumM$, which stores the number of passengers getting on to (off from) a journey in each stop but cannot track  
individual trips. On the other hand, in  $\tctr$, queries about the load of the transportation network,   such as \textit{ number of passengers into the vehicle in journey $j$ between stops $X$ and $X$$+$$1$},  would become very time consuming. 

\section{Experimental evaluation} \label{sec:experiments}



We created a synthetic collection of user trips generated from a GTFS\footnote{\url{https://developers.google.com/transit/gtfs/reference}} description of urban and medium-distance buses\footnote{Provided by CRTM (\url{http://www.crtm.es})} from  Madrid, with 1049 different lines and 10913 stop locations. 
We used real stop times from the journeys provided by the GTFS to generate ten million user trips over the span of a month. 
Each user trip created had one or more stages, defined as pairs $(stop\_in, stop\_out)$, being $stop\_in$ and $stop\_out$ respectively triples  $(stop, line, journey)$ that determine the stop, line, and the journey where the user got on and got off.

The created user trips started from a random stop on a random journey, and followed the stops on that journey. After at least two traversed stops, we used a probability table to determine if the stage ends and user switch lines. In such case, we simulate the user getting off from that journey and either waiting on the same stop (at most $30$ minutes) or walking to a nearby (100 meters) stop to get on to a new journey. 
We  ensured there were no inconsistencies in our generated trips (i.e. users getting on the same line from which they just got off). Trip lengths were limited to 100 stops. Yet, after each traversed stop, the probability for ending the trip  was $0.01\lambda$, where $\lambda$ is the 
number of previously traversed stops.

Finally, we represented all those trips using $\tctr$ and $\acumM$.

\subsubsection{Implementation details:}
\label{sec:implementation}

Due to the relatively small size of the network, the common structures were built using plain arrays of fixed size integers. Table~\ref{tab:common} shows the space occupied by these structures.


For $\tctr$, we used the $\csa$ from \cite{FBNCPR12} tuned the  sampling rate for $\Psi$ ($t_{\Psi}$) to the values $t_{\Psi} = \{32, 128, 512\}$. To represent bitvector $D$ we used a {\em SDArray} \cite{okanohara2007practical}.
In Table~\ref{tab:tctr_space} (left) we show the space required by $\Psi$, $D$, $B$, and $V'$ for an input of 35,702,981 entries in $S$ and $Jcodes$, when compared to a baseline that uses fixed width integers to represent the pairs $(line,stop)$ in the trips (that is, of $\lceil \log_2 |V| \rceil$ bits/entry, where $V$ is the vocabulary defined in Section~\ref{sec:csa}). 
In the $\wm$ of the $\tctr$ we used a $RRR$ bitvector to compress the bitmaps \cite{Raman:2002:SID:545381.545411}. We set  sampling parameter $s \in \{32,64,128\}$, as shown in the Table~\ref{tab:tctr_space} (right), where we compare the space of $\wm$ with a plain representation of the $journey$-$id$s that uses the number of bits needed to represent the maximum number of journeys on any line, $\operatorname*{arg\,max}_{l \in L} \lceil \log_2 J^l \rceil$.


\begin{table}[htbp]
	\begin{center}
		\scriptsize
		\begin{tabular}{|r|*{4}{c|}}
			\cline{2-5}
			\multicolumn{1}{c|}{} & {\em (i) Lines~~~~} & {\em (ii) Schedules~~~~} & {\em (iii) Inverted index of stops~~~~}  & Overall size\\
			\hline
			Size (KiB) & 176.54 & 7299.66 & 139.96 & 7616.17\\
			\hline
		\end{tabular}
        \normalsize
		\caption{Size of common structures (in KiB).}
		\label{tab:common}
	\end{center}
\end{table}

\begin{table}[htbp]
	\centering
	\setlength\tabcolsep{4pt}
	\begin{minipage}{0.48\textwidth}
		\centering
        \scriptsize
		\begin{tabular}{|l|*{3}{p{12mm}}|}
			\cline{2-4}
			\multicolumn{1}{c|}{} & \multicolumn{3}{c|}{$t_{\Psi}$ } \\
			\multicolumn{1}{c|}{} & 32 & 128 & 512 \\
			\hline
            $\icsa$ & 27.13 MB (39.84)\% & 21.91 MB (32.17)\% & 20.58 MB (30.22)\% \\
			\hline
		\end{tabular} 
	\end{minipage}%
	\hfill
	\begin{minipage}{0.48\textwidth}
		\centering
        \scriptsize
		\begin{tabular}{|l|*{3}{p{12mm}}|}
			\cline{2-4}
			\multicolumn{1}{c|}{} & \multicolumn{3}{c|}{RRR sampling } \\
			\multicolumn{1}{c|}{} & 32 & 64 & 128 \\
			\hline
            $\wm$ & 42.35 MB (71.07)\% & 39.43 MB (66.18)\% & 37.98 MB (63.73)\% \\
			\hline
		\end{tabular}
	\end{minipage}
	\caption{Compression of the $\csa$ (left) and $\wm$ (right) components from $\tctr$. Percentages refer to the sizes  with respect to the size of the uncompressed baseline.}	\label{tab:tctr_space}
\end{table}

\begin{table}[htbp]
	\begin{center}
		\scriptsize
		\begin{tabular}{|l|*{2}{r}|}
			\cline{2-3}
			\multicolumn{1}{r|}{} & Accumulated matrix & ~~~~Differential Matrix \\
			\hline
			get-on matrix & 11189 KB (100\%) & 5596 KB (50.01\%) \\
			get-off matrix & 11189 KB (100\%) & 5596 KB (50.01\%) \\
			\hline
		\end{tabular}
        \normalsize
		\caption{Sizes of the different $\acumM$ variants.}
		\label{tab:acum_space}
	\end{center}
\end{table}

For $\acumM$, we consider both the simple {\em accumulated-matrix} and the version using differential encoding. In the former case we have a simple matrix of integers. In the latter one, the middle column is kept apart as an integer array. For the rest of columns in the {\em differential matrix} values in each cell are encoded with  $\operatorname*\lceil \log_2 N \rceil$ bits, being $N$ the maximum difference (i.e. value on those columns). 

\subsubsection{Query execution times:}
We run experiments to show the query execution times of our proposals. An Intel Xeon E5-2620v4@2.1GHz machine was used.

On $\tctr$, we tested several configurations for the query {\em number of user trips from stop $X$ to stop $Y$}, labeled as $\mathsf{xy_{*}}$ in Table~\ref{tab:tctr_times}. The entry for $\mathsf{xy}$ with no subindices applies no line nor time restriction. $\mathsf{xy_{S}}$ and $\mathsf{xy_{E}}$ restrict, respectively, the \underline{S}tarting and \underline{E}nding stop to a specific line. $\mathsf{xy_{T}}$ denotes a \underline{T}emporal restriction (at one random day). Therefore, combinations of these subindices stand for combinations of these three restrictions. We randomly generated a set of $10,000$ query patterns by choosing trips from all the  available user trips. In the densest setup ($t_{\Psi}=32$, RRR = $32$) all the queries are answered in around $6$-$30 \mu$s.

The last row also includes the times to solve the query: {\em How many users got on in a stop $X$ on a given line during a given day?} ($\mathsf{J^kS^1}$). We also implemented this query in $\acumM$ to  compare the efficiency for these type of queries.

\begin{table}[htbp]
	\begin{center}
		\scriptsize
		\begin{tabular}{|l|*{9}{r}|}
			\cline{2-10}
			\multicolumn{1}{c|}{} & \multicolumn{3}{c|}{$RRR=32$} & \multicolumn{3}{c|}{$RRR=64$} & \multicolumn{3}{c|}{$RRR=128$} \\
			\multicolumn{1}{c|}{} & $t_{\Psi}=32$ & $t_{\Psi}=128$ & $t_{\Psi}=512$ & $t_{\Psi}=32$ & $t_{\Psi}=128$ & $t_{\Psi}=512$ & $t_{\Psi}=32$ & $t_{\Psi}=128$ & $t_{\Psi}=512$ \\
            \hline
              $\mathsf{xy      }$ & 6.94 & 10.28 & 22.89 & 6.90 & 10.23 & 22.80 & 6.87 & 10.21 & 22.82 \\
              $\mathsf{xy_{S}  }$ & 6.89 & 10.23 & 22.95 & 6.86 & 10.22 & 22.82 & 6.90 & 10.21 & 22.74 \\
              $\mathsf{xy_{E}  }$ & 29.37 & 62.63 & 192.86 & 29.13 & 62.60 & 192.01 & 29.14 & 62.38 & 192.28 \\
              $\mathsf{xy_{SE} }$ & 29.21 & 61.84 & 192.11 & 28.88 & 61.83 & 190.46 & 28.92 & 62.05 & 190.47 \\
              $\mathsf{xy_{T}  }$ & 31.85 & 64.66 & 195.62 & 31.63 & 65.52 & 195.06 & 31.95 & 65.47 & 195.31 \\
              $\mathsf{xy_{ST} }$ & 31.61 & 63.83 & 193.41 & 31.12 & 65.05 & 193.14 & 31.54 & 64.78 & 194.23 \\
              $\mathsf{xy_{ET} }$ & 31.73 & 65.04 & 195.53 & 31.66 & 65.73 & 195.15 & 32.01 & 65.09 & 195.45 \\
              $\mathsf{xy_{SET}}$ & 31.36 & 63.96 & 194.12 & 31.17 & 64.99 & 193.11 & 31.42 & 64.42 & 193.40 \\
			\hline
			$\mathsf{J^kS^1}$ & 5.05 & 5.85 & 9.14 & 5.20 & 6.01 & 9.30 & 5.41 & 6.31 & 9.63 \\
			\hline
		\end{tabular}
        \normalsize
		\caption{Performance at query time shown in $\mu$secs/query for $\tctr$.}
		\label{tab:tctr_times}
	\end{center}
\end{table}



\begin{table}[htbp]
	\begin{center}
		\scriptsize
		\begin{tabular}{|l|*{2}{l}|}
			\cline{2-3}
			\multicolumn{1}{c|}{}  & Accumulated matrix~~~~ & Differential Matrix \\
			\hline
			$\mathsf{J^kS^1}$(column)    & 131  & 211 \\
			\hline
			$\mathsf{J^1S^*}$ (rows)  & 107 & 221 \\			
			$\mathsf{J^kS^k}$ (window) & 76 & 182 \\			
			\hline
		\end{tabular}
        \normalsize
		\caption{Performance at query time for the variants of $\acumM$ (in $ns$ per query).}
		\label{tab:acum_time}
	\end{center}
\end{table}


To test $\acumM$ we considered three types of queries: The query $\mathsf{J^kS^1}$ discussed above; {\em total number of passengers that got on in all the stops of a given journey (1-row)}, labeled $\mathsf{J^1S^*}$; and {\em total passengers that got on along a range of consecutive stops from several consecutive journeys (window)}, labeled $\mathsf{J^kS^k}$. We generated $20,000$ query patterns based on the real data (line number, stop number, and journeys), and then run the tests obtaining query times around $0.1 \mu$s when using the accumulated matrix. As expected, the differential accumulated matrix performs around twice slower. Yet, $\acumM$ performs more than one order of magnitude faster than $\tctr$ on query $\mathsf{J^kS^1}$. Table~\ref{tab:acum_time} shows the results.


\section{Conclusions and future work} \label{sec:conclusions}

We have analyzed the problem of representing trips over a public transportation network and presented two data structures designed to efficiently answer two subsets of useful queries. Both approaches use some common data structures defined to represent the transportation network, that is the \textit{offer} (lines, schedule of their journeys and stops) it provides.

The first  proposal, $\tctr$, represents the whole set of user trips during a period of time. Each user trip is composed of stages performed over specific journeys of different lines. This data structure is useful to analyze user trip patterns, that represent the real \textit{demand} over the transportation network. Yet, $\tctr$ enables not only counting-based queries for the number of passengers related to any stop of the network, but also queries for stops or stops-lines were users start/end trips or switch lines. It also allows to retrieve individual trips.

The second structure ($\acumM$) focuses on the usage of lines. For each line, it keeps in an accumulated fashion, the number of passengers that, at each stop, get on/off from the vehicle performing a specific journey of the line. This simplifies solving queries regarding the load of the different journeys and therefore to analyze when specific lines must be reinforced with a more frequent schedule.

We understand  more research is needed in this topic. Even we can see $\acumM$ as a data warehouse were we store basically the same information but in an aggregated way, we consider that avoiding the redundancy between both structures would be desirable. As future work, we are interested in developing a unique self-indexed structure providing the functionality included in both $\tctr$ and $acumM$.

\bibliographystyle{splncs04}
\bibliography{refs}

\begin{thebibliography}{10}
\providecommand{\url}[1]{\texttt{#1}}
\providecommand{\urlprefix}{URL }
\providecommand{\doi}[1]{https://doi.org/#1}

\bibitem{BRISABOA20181}
Brisaboa, N.R., Fari{\~n}a, A., Galaktionov, D., Rodriguez, M.A.: A compact
  representation for trips over networks built on self-indexes. Information
  Systems  \textbf{78},  1--22 (2018). \doi{10.1016/j.is.2018.06.010}

\bibitem{CNO15}
Claude, F., Navarro, G., {Ord{\'o\~n}ez}, A.: The wavelet matrix: An efficient
  wavelet tree for large alphabets. Information Systems  \textbf{47},  15--32
  (2015)

\bibitem{FBNCPR12}
Fari{\~n}a, A., Brisaboa, N.R., Navarro, G., Claude, F., Places, {\'A}.S.,
  Rodr\'{\i}guez, E.: {Word-based Self-Indexes for Natural Language Text}. ACM
  Transactions on Information Systems  \textbf{30}(1),  article 1: (2012).
  \doi{10.1145/2094072.2094073}

\bibitem{DBLP:conf/focs/FerraginaM00}
Ferragina, P., Manzini, G.: Opportunistic data structures with applications.
  In: Proc. 41st IEEE Symposium on Foundations of Computer Science (FOCS). pp.
  390--398 (2000)

\bibitem{DBLP:conf/w2gis/FunkeSSS15}
Funke, S., Schirrmeister, R., Skilevic, S., Storandt, S.: Compass-based
  navigation in street networks. In: Proc. 14th International Symposium on Web
  and Wireless Geographical Information Systems (W2GIS). pp. 71--88. LNCS 9080
  (2015)

\bibitem{DBLP:journals/jss/KellarisPT13}
Kellaris, G., Pelekis, N., Theodoridis, Y.: {Map-matched Trajectory
  Compression}. Journal of Systems and Software  \textbf{86}(6),  1566--1579
  (2013). \doi{10.1016/j.jss.2013.01.071}

\bibitem{DBLP:conf/gis/KoideTY15}
Koide, S., Tadokoro, Y., Yoshimura, T.: {SNT-index}: Spatio-temporal index for
  vehicular trajectories on a road network based on substring matching. In:
  Proc. 1st International ACM SIGSPATIAL Workshop on Smart Cities and Urban
  Analytics ({UrbanGIS@SIGSPATIAL}). pp.~1--8 (2015).
  \doi{10.1145/2835022.2835023}

\bibitem{DBLP:conf/gis/KroghPTT14}
Krogh, B., Pelekis, N., Theodoridis, Y., Torp, K.: Path-based queries on
  trajectory data. In: Proc. 22nd {ACM} {SIGSPATIAL} International Conference
  on Advances in Geographic Information Systems ({SIGSPATIAL)}. pp. 341--350
  (2014)

\bibitem{Munizaga20129}
Munizaga, M.A., Palma, C.: {Estimation of a disaggregate multimodal public
  transport Origin–Destination matrix from passive smartcard data from
  Santiago, Chile}. Transportation Research Part C: Emerging Technologies
  \textbf{24},  9--18 (2012). \doi{10.1016/j.trc.2012.01.007}

\bibitem{okanohara2007practical}
Okanohara, D., Sadakane, K.: Practical entropy-compressed rank/select
  dictionary. In: Proc. 9th Workshop on Algorithm Engineering and Experiments
  (ALENEX). pp. 60--70 (2007). \doi{10.1137/1.9781611972870.6}

\bibitem{Raman:2002:SID:545381.545411}
Raman, R., Raman, V., Rao, S.S.: Succinct indexable dictionaries with
  applications to encoding k-ary trees and multisets. In: Proc. 13th Annual
  ACM-SIAM Symposium on Discrete Algorithms (SODA). pp. 233--242 (2002)

\bibitem{DBLP:journals/josis/RichterSL12}
Richter, K.F., Schmid, F., Laube, P.: {Semantic Trajectory Compression:
  Representing Urban Movement in a Nutshell}. Journal of Spatial Information
  Science  \textbf{4}(1),  3--30 (2012). \doi{10.5311/JOSIS.2012.4.62}

\bibitem{Sad03}
Sadakane, K.: New text indexing functionalities of the compressed suffix
  arrays. Journal of Algorithms  \textbf{48}(2),  294--313 (2003)

\end{thebibliography}

\end{document}